\documentstyle[prc,epsf,aps]{revtex}

\begin{document}
%\draft
\title{Polarization observables in electronuclear two-nucleon
knockout} 
\author{Jan Ryckebusch \footnote{E-mail :
jan.ryckebusch@rug.ac.be}, Wim Van Nespen and Dimitri Debruyne}
\address{Department of Subatomic and Radiation Physics \protect\\
University of Gent, Proeftuinstraat 86, B-9000 Gent, Belgium}

\date{\today}
\maketitle
\begin{abstract}
Differential (e,e$'$pp) measurements are presently recognized as a way
of studying short-range correlations in finite nuclei.  The
($\vec{\mathrm{e}}$,e$'\vec{\mathrm{p}}$p)  and
($\vec{\mathrm{e}}$,e$'\vec{\mathrm{p}}$n) differential cross section
and polarization observables are studied in a microscopic model that
accounts for the short-range correlations, outgoing-nucleon
distortions, meson-exchange and $\Delta$-isobar currents. It is
pointed out that polarization observables represent an attractive
alternative for absolute electronuclear two-nucleon knockout
measurements. In the polarization transfer P$'_t$ for 
($\vec{\mathrm{e}}$,e$'\vec{\mathrm{p}}$p), the
effect of central short-range correlations is predicted to be large
while at the same time the final-state interaction effects are small.

%for the
%$^{16}$O($\vec{\mathrm{e}}$,e$'\vec{\mathrm{p}}$p)4 and the
%$^{16}$O($\vec{\mathrm{e}}$,e$'\vec{\mathrm{p}}$n) reaction to
%specific states in $^{14}$C and $^{14}$N as well as
%$^{12}$C($\vec{\mathrm{e}}$,e$'\vec{\mathrm{p}}$p) and
%$^{12}$C($\vec{\mathrm{e}}$,e$'\vec{\mathrm{p}}$n) for the complete
%p-shell.
\end{abstract}



%24.70.+s Polarization phenomena in reactions
%25.30.-c Lepton-induced reactions
%24.10.-i Nuclear-reaction models and methods

For long, polarization degrees of freedom have been advocated as an
attractive way of obtaining nucleon and nuclear structure information
from exclusive electron scattering experiments.  Often, polarization
observables allow for accessing very specific combinations of
structure functions with transparent physical interpretations.
Recently explored examples include the precise measurement of the
neutron electromagnetic form factors for which extensive programs were
set up at the intermediate-energy electron accelerator facilities 
MAMI, BATES and TJNAF.  In these neutron form factor
studies, polarized electron beams are used in combination with either
a polarized target ($^3 \vec{\mathrm He}$($\vec{\mathrm e}$,e$'$)) 
or an ejectile
polarimeter (d($\vec{\mathrm e}$,e$' \vec{\mathrm n}$)p).

  Although various electronuclear reactions, like the semi-inclusive
eA$\longrightarrow$ e$'$pX \cite{benhar97} and the inclusive (e,e$'$)
\cite{benhar94,ciofi94}, have been shown to exhibit some sensitivity
to the short-range structure of finite nuclei, the triple-coincidence
(e,e$'$pp) reaction, albeit a rather involving coincidence reaction,
is probably one of the most direct ways of probing the two-nucleon
correlations \cite{laget}.  Not only does the electromagnetic interaction avoid
complications with respect to the initial-state interaction, the
measured properties of the two detected protons makes it possible to
reconstruct the characteristics of the correlated diproton in the
target nucleus.  Recently, it was experimentally verified
\cite{blom98,guenther,gerco} that even at relatively low values of the
four-momentum transfer ($Q^2 \leq$ 0.12~$GeV^2$) and selected
kinematics the (e,e$'$pp) reaction is dominated by photoabsorption on
$^1S_0$ diprotons \cite{guenther,gerco}.  This observation confirms
the potential of the exclusive (e,e$'$pp) reaction to probe the
short-range part of the two-nucleon correlations in the nuclear
medium.  At the same time, it is now widely recognized that
electromagnetic production of a $\Delta _{33}$ resonance with
subsequent $\pi ^0$ decay and reabsorption on a proton represents an
important competing mechanism that also feeds the (e,e$'$pp) reaction.
Accordingly, it is of the utmost importance to find observables that
allow for an unambiguous determination of the short-range correlation
effects out of the background of competing processes. As the
(e,e$'$pp) strength generated through intermediate $\Delta _{33}$
creation is predominantly transverse, variables that depend strongly
on the longitudinal channel are the obvious candidate to meet this
condition.  Here, we investigate the potential of polarization
observables to gain a better quantitative control on the different
reaction mechanisms contributing to direct electronuclear two-nucleon
knockout.  With the eye on future experimental investigations of
proton-neutron correlations at TJNAF and MAMI, we also address the
(e,e$'$pn) channel in our investigations.

\begin{figure}
\begin{center}
{\mbox {\epsfxsize=15.cm \epsffile{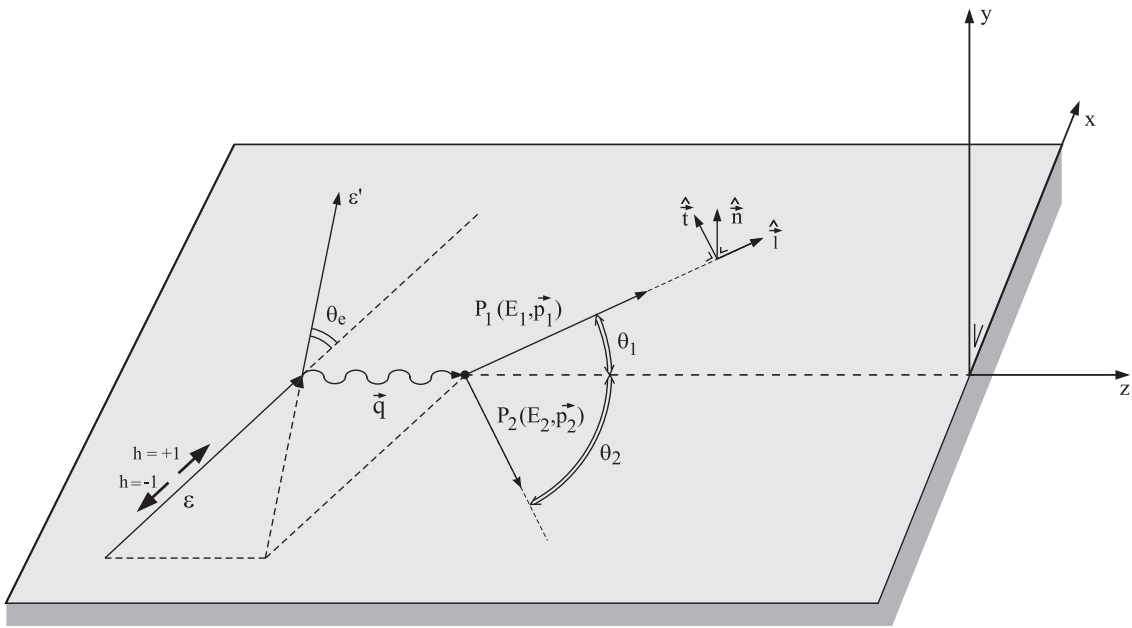}}}
\end{center}
\caption{The coordinate system for the coplanar
A($\vec{\mathrm{e}}$,e$'\vec{\mathrm{p}}$N) reaction in the laboratory
frame.  The recoil polarimetry is performed on the nucleon
characterized by the four-vector $P_1$.}
\label{fig:kinemat}
\end{figure}

The eightfold A($\vec{\mathrm{e}}$,e$'$N$_1$N$_2$) 
differential cross section
for excitation of a narrow discrete state in the residual nucleus
reads
\begin{eqnarray}
& & {d^8 \sigma \over dE_1 d \Omega _1 d \Omega _2 d \epsilon ' d \Omega
_{\epsilon '}} ({\mathrm \vec{e},e'N_1N_2}) = 
{1 \over 4 (2\pi)^8 } p_1 p_2 E_1 E_2 f_{rec}^{-1} \sigma_{M}
\nonumber \\ 
\times
& & \Biggl[ v_T W_T
+ v_L W_L
+ v_{LT} W_{LT}
+ v_{TT} W_{TT} + h \biggl[ v'_{LT} W'_{LT} + v'_{TT} W'_{TT} \biggr] \Biggr]
\label{eq:eepnn}
\end{eqnarray}
where $h$=$\pm$1 is the incoming electron helicity, $f_{rec}$ a recoil
factor and $\sigma _M$ the Mott cross section.  The functions
$v_{\mathrm{i}}$ (i=L,T,LT,TT) and $v'_{\mathrm{i}}$ (i=LT,TT) contain
the electron kinematics.  They are defined along the conventions of
Ref.~\cite{raskin}. The polar and azimuthal angles are determined
relative to the electron scattering plane which is conventionally
taken as the $xz$ plane (Figure~\ref{fig:kinemat}). 
The above differential cross section is of the familiar form
\begin{equation}
{d^8 \sigma \over dE_1 d \Omega _1 d \Omega _2 d \epsilon ' d \Omega
_{\epsilon '}} (\mathrm{\vec{e},e'N_1N_2}) = \sigma _0 ( 1 + h A) \; ,
\end{equation}
with A the electron analyzing power and $\sigma _0$ the unpolarized
cross section.  All of the structure functions $W$ in the cross
section of Eq.~(\ref{eq:eepnn}) depend on the variables
(q,$\omega$,p$_1$,p$_2$,$\theta _1$,$\theta_2$ and $\phi _1 - \phi
_2$) in a non-trivial manner.  Three terms (W$_{LT}$, W$_{TT}$ and
W$'_{LT}$) have an additional azimuthal dependency on the variable $
\phi _1 + \phi_2 \over 2$ that can be pulled out of the structure
functions \cite{boffigiusti,donnelly}.  In doing this, each of the
structure functions W$_{LT}$, W$_{TT}$ and W$'_{LT}$ falls apart in
two terms \cite{giustifirst}.  In experiments in which none of the
polarizations of the produced hadrons are determined, the W$'_{TT}$
vanishes in all kinematical situations and the W$'_{LT}$ in coplanar
kinematics.

Here, we also consider the situation that the polarization of one
ejectile is determined.  In that particular case, one
obtains a differential cross section in the standard form
\begin{equation}
{d^8 \sigma \over dE_1 d \Omega _1 d \Omega _2 d \epsilon ' d \Omega
_{\epsilon '}} (\mathrm{\vec{e},e'\vec{N}_1N_2})
= \frac {\sigma _ 0} {2} \left[1 +  \vec{P} \cdot \vec{\sigma} + h ( A
+ \vec{P}' \cdot \vec{\sigma}) \right] \; ,
\end{equation}
here $\vec{\mathrm P}$ is the induced polarization and $\vec{\mathrm
P}'$ the polarization transfer.  As is commonly done, the polarization
of the escaping hadron is expressed in terms of the reference frame
determined by the following unit vectors
\begin{equation}
\hat{\vec{l}} = \frac {\vec{p}_1} {\left| \vec{p}_1 \right|}
\; \; \; \; \;
\hat{\vec{n}} = \frac {\vec{q} \times \vec{p}_1} 
{\left| \vec{q} \times \vec{p}_1 \right|}
\; \; \; \; \;
\hat{\vec{t}} = \hat{\vec{n}} \times \hat{\vec{l}} \; 
\end{equation} 
with $\hat{\vec{l}}$ pointing in the direction of the momentum
$\vec{p}_1$ of the proton for which the recoil polarization is
determined.  In the coplanar case, $\hat{\vec{t}}$ is in the electron
scattering plane and $\hat{\vec{n}}$ is perpendicular to it.  This
situation is illustrated in Figure~\ref{fig:kinemat}. The experimental
determination of the induced
polarization and polarization transfer along one of these directions
($\vec{\mathrm P} _i$, $\vec{\mathrm P}' _i$ with $i$=(n,l,t)) 
is found by measuring the 
following ratios
\begin{eqnarray}    
{\mathrm P}_i & = &  \frac {\sigma (s_{1i}\uparrow) -  
\sigma (s_{1i}\downarrow)} 
{\sigma (s_{1i}\uparrow) + \sigma (s_{1i}\downarrow)} \nonumber \\
{\mathrm P}_i' & = &  \frac {
\left[ \sigma (h=+1,s_{1i}\uparrow) - \sigma (h=-1,s_{1i}\uparrow) \right] 
- \left[ \sigma (h=+1,s_{1i}\downarrow) - 
\sigma (h=-1,s_{1i}\downarrow) \right] }
{
\left[ \sigma (h=+1,s_{1i}\uparrow) + \sigma (h=-1,s_{1i}\uparrow) \right] 
+ \left[ \sigma (h=+1,s_{1i}\downarrow) +  
\sigma (h=-1,s_{1i}\downarrow) \right] }
\end{eqnarray}
where $s_{1i}\uparrow$ indicates that the proton labelled as ``1'' is
spin-polarized in the positive $\hat{\vec{i}}$-direction, and 
$s_{1i}\downarrow$ in the negative.

In a recent publication \cite{giusti98}, it was pointed out that the
structure of the finite nucleus can act as a filter for the different
mechanisms contributing to the (e,e$'$pp) reaction.  
%Some final states
%are predominantly feeded through the ground-state correlations,
%whereas for others the observed two-proton knockout strength is
%predicted to be exclusively the result of intermediate
%electroexcitation of a $\Delta$ with subsequent decay in two protons.
Elaborating on the suggestion that the structure of the final state
may be selective as far as the dominating reaction mechanisms is
concerned, we have considered the $^{16}$O(e,e$'$pp) reaction for
excitation of specific states in the discrete part of the energy
spectrum in the residual nucleus.  This reaction is currently under
investigation at NIKHEF \cite{gerco} and MAMI \cite{guenther}. The
numerical calculations of which the results are presented here are
performed within the model outlined in Ref.~\cite{jan97}.  It involves
a non-relativistic distorted wave calculation in a shell-model
framework.  The discrete final state of the residual nucleus is
conceived as a linear combination of two-hole states relative to the
ground-state of the target nucleus.  The distorted outgoing nucleon
waves are generated through a partial wave expansion in terms of the
continuum eigenfunctions of the mean-field potential.  As the latter
is also used to calculate the bound-state wave functions the
orthogonality between initial and final states is obeyed thus avoiding
that spurious effects contribute to the calculated cross sections. 
This procedure neglects the final-state interaction between the
escaping nucleons.  
The
$\Delta p \longrightarrow pp$ decay is assumed to be regulated by $\pi
^0$ decay.  In constructing the two-body isobaric current that
corresponds with this process, the standard $\pi NN$, $\pi N \Delta$
and $\gamma N \Delta$ couplings are introduced \cite{jan97}.  
The $\Delta$ propagators that are used in the present work are
described in Ref. ~\cite{jan97}.  Apart from the standard $\pi N$ decay
width, an extra 40 MeV complex width is introduced.  The latter
accounts for the medium modifications and was found necessary in order
to reach a fair agreement between the model predictions and the
$^{12}$C($\gamma$,pp) and $^{12}$C($\gamma$,pn) data in the
$\Delta$-resonance region \cite{douglas}.
The
ground- and final-state correlations are implemented through the
introduction of a central Jastrow correlation function $g(r_{12})$.
In calculating the transition between the (correlated) initial and
(correlated) final state we adopt a cluster expansion in terms of the
correlation function retaining these terms with a single correlation
line that is directly connected to the two ejectiles. This procedure
is equivalent with a lowest-order cluster expansion in which all
three-point diagrams have been neglected \cite{jan97,giam}.  Recent
calculations addressed the amount of two-nucleon knockout strength
induced by the ground-state correlations to the inclusive longitudinal
$^{12}$C(e,e$'$) structure function starting from a correlated
groundstate and evaluating both the two-point and three-point
diagrams.  The latter were stressed to be essential in order to
guarantee the wave-function normalization in an inclusive (e,e$'$)
calculation.  Nevertheless, the two-point diagrams were found to
represent the major source of 2N knockout strength.  The 2N knockout
strength from the three-point diagrams was found to be considerably
smaller and to simply sum up to the strength from the two-point
diagrams.  Here, we restrict ourselves to {\em exclusive} (e,e$'$NN)
reactions in kinematics that favour two-nucleon processes.  The
normalization of the final A-body wave functions is fixed through its
asymptotic behaviour.  All this justifies the neglect of the
three-point diagrams that would rather play a role in kinematic
regions where three-nucleon kinematics is favoured. Unless otherwise
specified the results of this paper are obtained with the central
correlation function from Ref.~\cite{gearhart} that is obtained from a
G-matrix calculation.

\begin{figure}
\begin{center}
\setlength{\unitlength}{1cm}
\begin{picture}(16,12)
\put(-1.00,-4.0){\mbox{\epsfysize=15.cm\epsffile{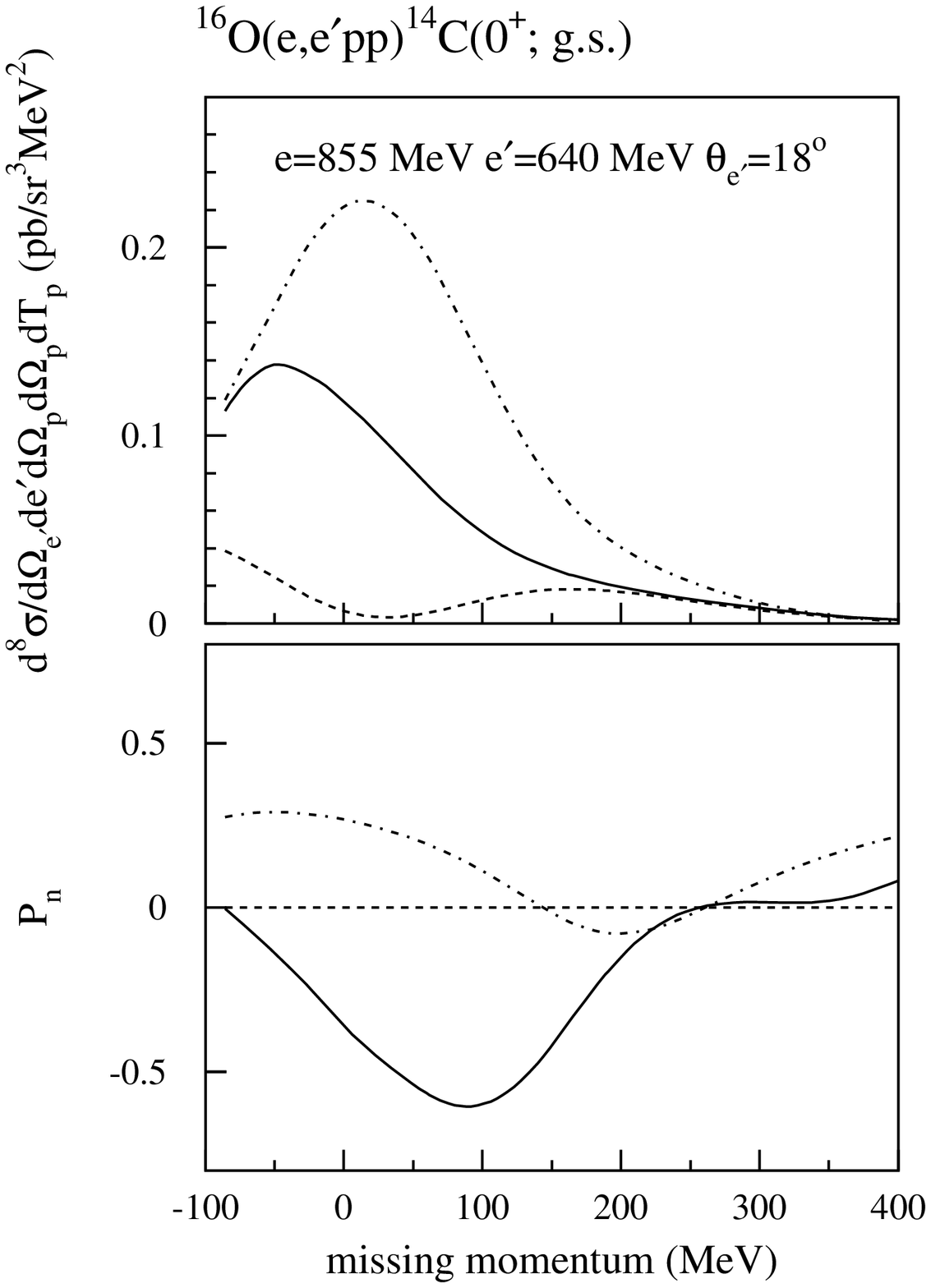}}}
\put(6.50,-4.0){\mbox{\epsfysize=15.cm\epsffile{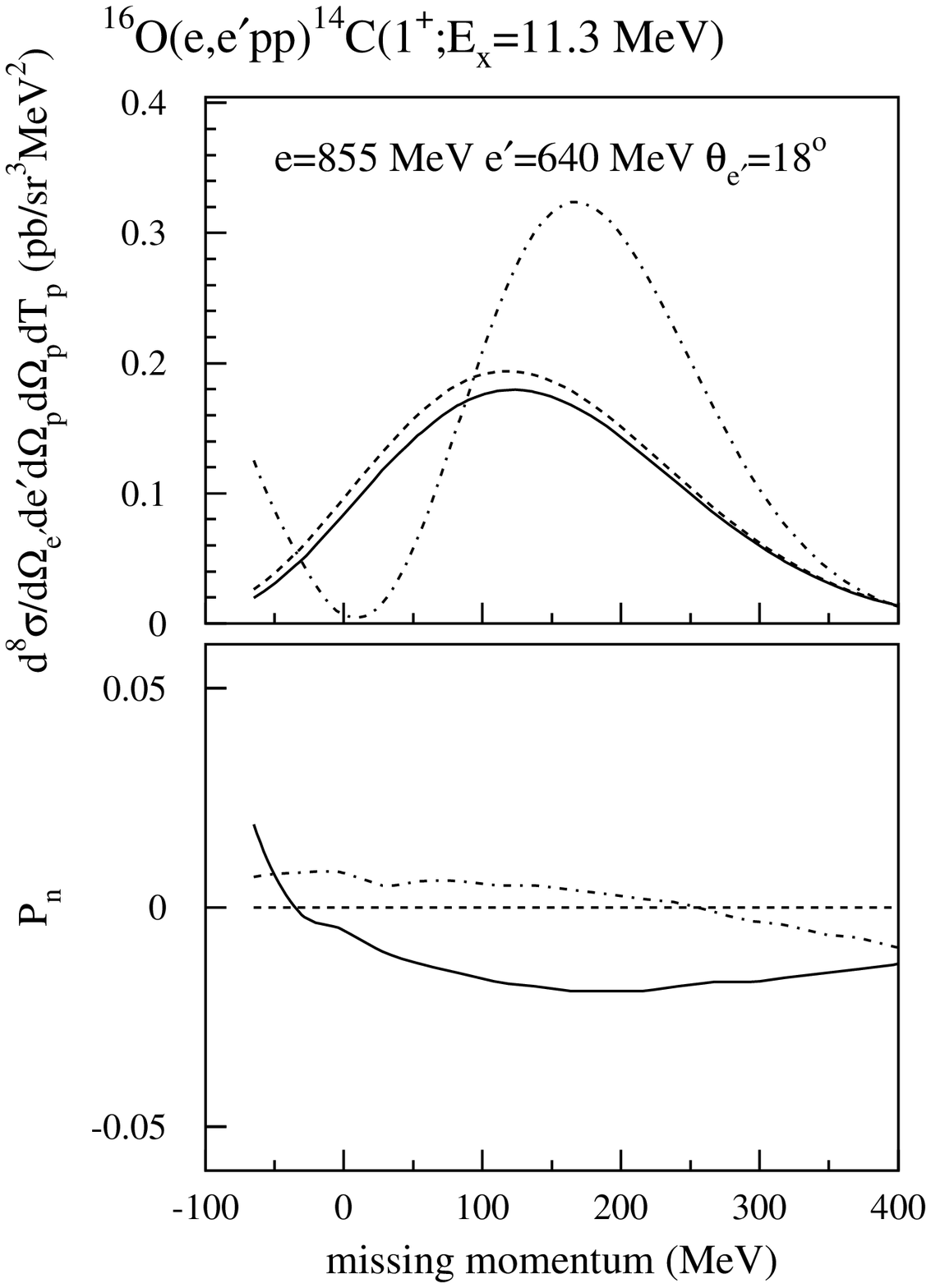}}}
\end{picture}
\end{center}
\caption{The missing momentum dependence of the $^{16}$O(e,e$'$pp)
differential cross section and polarization observables in
superparallel kinematics. The solid curve is calculated in the
distorted-wave approximation including the $\Delta$-current and
ground-state correlations.  The latter are implemented through the
Jastrow function from Ref.~\protect \cite{gearhart}. The dot-dashed
curve is the equivalent of the solid line but is calculated with
plane wave outgoing nucleons waves.  The dashed line is the result of
a distorted-wave calculation including only the $\Delta$ current.}
\label{fig:eeppres}
\end{figure}

The specific kinematic situation in which the two hadrons are detected
along the direction of the three-momentum transfer $\vec{q}$ is
referred to as ``super-parallel'' kinematics.  As the resonant terms
in the elementary pion photoproduction operator are transverse in
nature \cite{osterfeld,lee97} one expects that contributions from
$\gamma^* pp \longrightarrow \Delta ^+ p \longrightarrow p p \pi^0
\longrightarrow pp $ to (e,e$'$pp) will be suppressed in ``super-parallel''
kinematics.  In this particular case, only two structure functions
($W_L$ and $W_T$) will contribute to the differential cross section
and the polarization observables P$_n$, P$'_l$, P$'_t$ are uniquely
determined by the $W_{LT}$, $W'_{TT}$ and $W'_{LT}$ terms
respectively.  In that sense, super-parallel kinematics in
($\vec{\mathrm{e}}$,e$'\vec{\mathrm{p}}$p) resembles the selectivity
of the ($\vec{\mathrm{e}}$,e$'\vec{\mathrm{p}}$) situation in parallel
kinematics \cite{boffigiusti,kelly}.  The results of
Figure~\ref{fig:eeppres} are obtained in super-parallel kinematics for
electron energies and angles that coincide with the central values
adopted in the MAMI measurements \cite{guenther}.  The cross sections
and polarization observables of Fig.~\ref{fig:eeppres} are shown as a
function of the missing momentum.  In super-parallel kinematics, this
variable is determined by $\left| \vec{P} \right| = - \left| \vec{p}_1
\right| + \left| \vec{p}_2 \right| + \left| \vec{q} \right|$, where we
have adopted the convention that $\vec{p}_1$ is parallel and
$\vec{p}_2$ anti-parallel to $\vec{q}$.  The variation in missing
momentum is reached by varying the kinetic energy of the nucleon
escaping in the direction of the three-momentum transfer
$\vec{q}$. Negative missing momenta correspond with a fast forward and
a slow backward proton.  As the missing momentum gets larger the
backward going nucleon (anti-parallel to $\vec{q}$) gains in kinetic
energy.  For the nuclear structure inputs in our calculations we rely
on the results of a recent calculation \cite{geurts} that predicted
the following two-hole structure for the $^{14}$C wave functions
$\left| 0^+ ; g.s. \right> = 0.77 \left| \left( 1p_{1/2} \right)^{-2}
; 0^+ \right> + 0.18 \left| \left( 1p_{3/2} \right)^{-2} ; 0^+ \right>
$ and $\left| 1^+ ; E_{x}=11.3~\mathrm{MeV} \right> = 0.77 \left|
\left( 1p_{1/2} \right)^{-1} \left( 1p_{3/2} \right)^{-1}; 1^+
\right>$.  
A recent analysis of the $^{16}$O(e,e$'$pp) data \cite{gerco} provided
some evidence for the realistic character of these wave functions.
More detailed information about the different wave function amplitudes
will become available from high-resolution experiments that are
presently under analysis \cite{guenther,gerco}.
The curves of Figure~\ref{fig:eeppres} confirm that the
structure of the final state is very selective with respect to the
reaction mechanism.  Whereas short-range correlations dominate the
ground-state transition their effect is hardly visible for the 1$^+$
transition that is completely dominated by intermediate $\Delta ^+$
creation and subsequent two-proton knockout.  The dominant reaction
mechanism reflects itself very clearly in the polarization
observable P$_n$ that is driven by  longitudinal-transverse
interference effects.
%Whereas the contribution
%stemming from the $\Delta$ produces an (induced) polarization which is
%varying slowly with the missing momentum, the short-range correlations
%that reflect themselves in longitudinal contributions, have a major
%effect on P$^n$, P$'^t$ and P$^l$.  
The predicted longitudinal excitation of the 1$^+$ state is so
marginal over the whole missing momentum range covered that the P$_n$
is almost vanishing.  For the 0$^+$ state, on the other hand, the
longitudinal strength produces a large normal component of the induced
polarization through interference of the central correlation effects
with the transverse background. Clearly, the P$_n$ is a measure for
the strength induced by the central short-range correlations.  Also
shown in Fig.~\ref{fig:eeppres} is the result of a plane-wave (PW)
calculation. This limiting case was reached by using spherical Bessel
functions in the partial wave expansions for the two ejectiles without
affecting the other ingredients of the calculations.  The PW
groundstate and 11.3~MeV missing momentum distribution look
respectively like an $F_{1S}$ and $F_{1P}$ center-of-mass distribution
for the initial pair.  Note that this is not such a trivial result as
one might conceive it, as the PW approximation is not a sufficient
condition to reach a formal factorization of the (e,e$'$pp) cross
section in terms of $F_{NL}$(P) ($NL$ denoting the quantum numbers
corresponding with the c.o.m. motion of the nucleon pair)
\cite{gottfried,janplb}.  The apparent scaling in terms of the
$F_{NL}$(P) in unfactorized calculations lends support for the
factorized approaches to electronuclear diproton knockout.  The major
effect of the distortion effects is a reduction of the cross section
and a shift in missing momentum.  This shift can be understood by
considering that there is substantial asymmetry in the kinetic
energies of the two ejectiles in super-parallel kinematics.

\begin{figure}
\begin{center}
\setlength{\unitlength}{1cm}
\begin{picture}(16,12)
\put(-1.00,-4.0){\mbox{\epsfysize=15.cm\epsffile{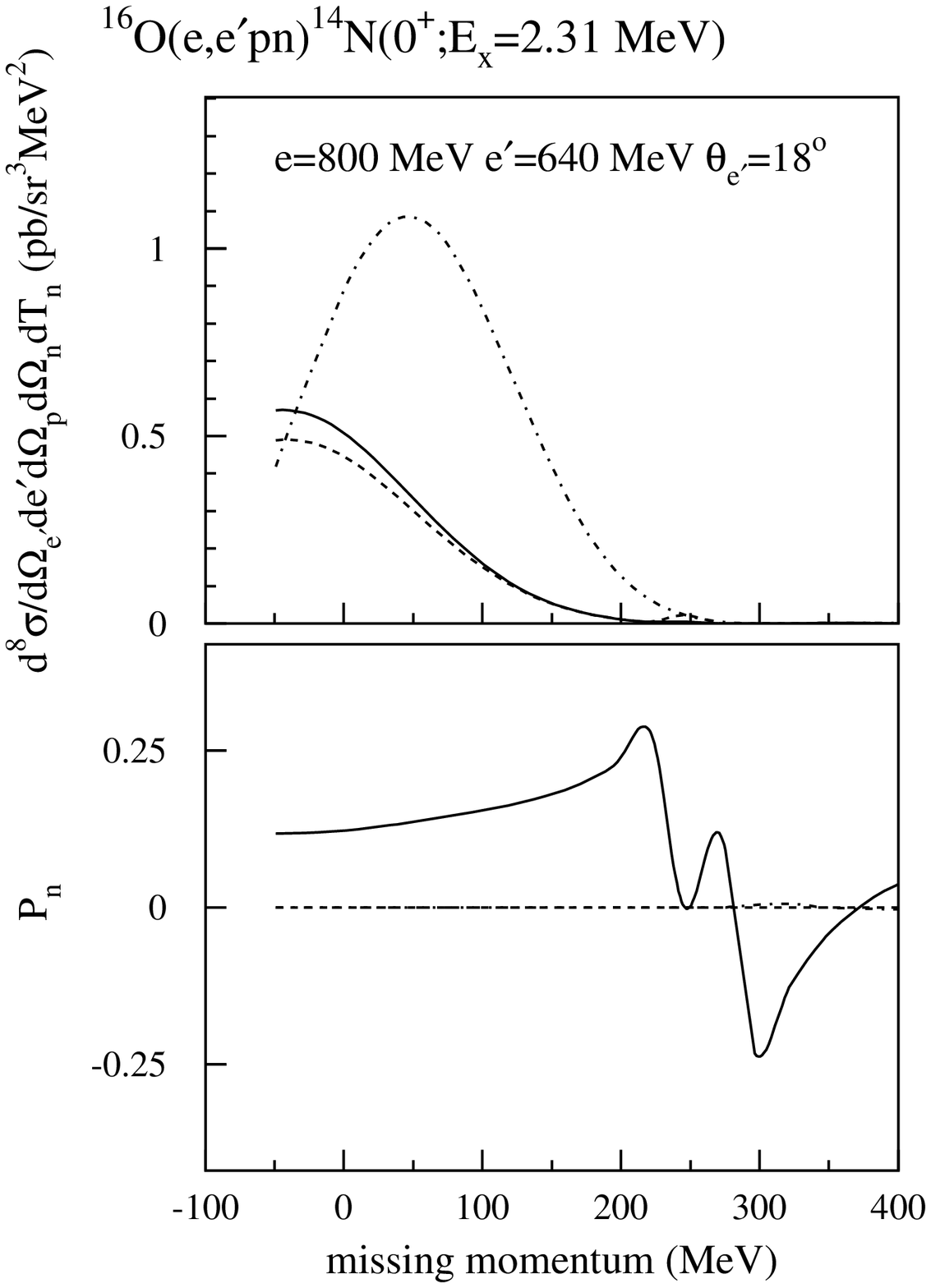}}}
\put(6.50,-4.0){\mbox{\epsfysize=15.cm\epsffile{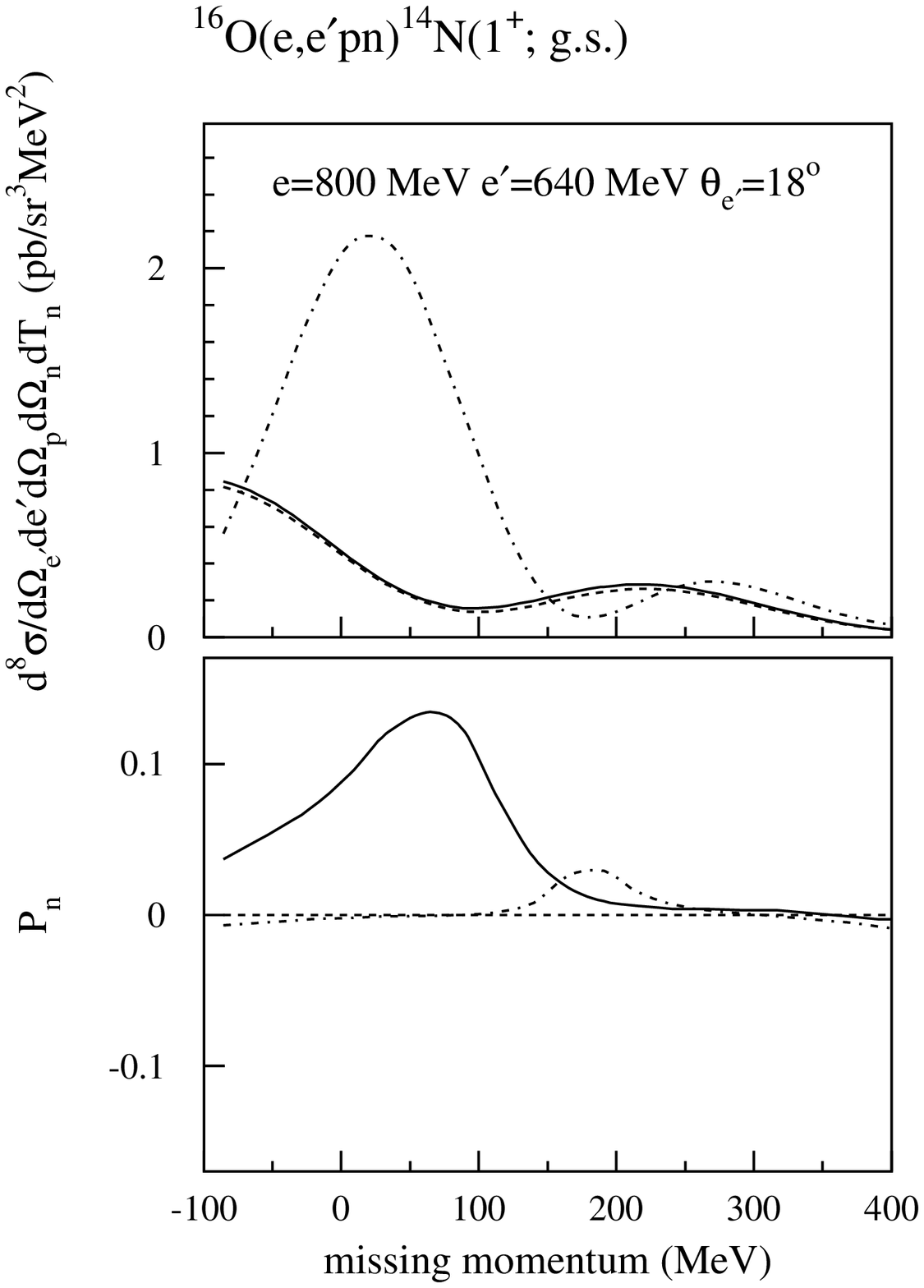}}}
\end{picture}
\end{center}
\caption{As in figure \ref{fig:eeppres} but now for the $^{16}$O(e,e$'$pn)
reaction.}
\label{fig:eepnres}
\end{figure}

The corresponding (e,e$'$pn) results for the two-proton knockout
results of Figure~\ref{fig:eeppres} are shown in
Figure~\ref{fig:eepnres}. We have considered a $(1p_{1/2})^{-2}$
shell-model structure for the lowest two states in $^{14}$N. In the
proton-neutron knockout calculations also the pion-exchange currents
are included.  This additional source of proton-neutron knockout
strength, which is transverse in nature, makes the relative
contribution of the central short-range correlations to the
differential cross section to be smaller.
The minor role of the central correlations relative to the
meson-exchange and isobaric contributions makes the $P_n$ to be
relatively small for proton-neutron knockout in super-parallel
kinematics.  The predicted missing momentum dependence for the 0$^+$
state bears a strong resemblance with the corresponding (e,e$'$pp)
results of Fig.~\ref{fig:eeppres} and has a clear S-wave shape,
pointing towards photoabsorption on proton-neutron pairs in a relative
$^1S_0$(T=1) state. Strong absorption on proton-neutron pairs in a
relative $^{2S+1} P_{J}$ state would imply a missing momentum
dependence for the 1$^+$ state that has a considerable $F_{1P}$
component.  This situation was encountered for the 1$^+$ state in the
proton-proton knockout results of Fig.~\ref{fig:eeppres}.  The
calculated (e,e$'$pn) missing momentum dependence, however, reflects a
$F_{NS}$(N=1,2) shape which for the considered shell-model
configuration points towards photoabsorption on $^3S_1(T=0)$ and
$^3D_1(T=0)$ proton-neutron pairs \cite{ryc98}.  This result confirms
the strong likelihood of the electromagnetic probe to couple to
``quasi-deuteron'' like proton-neutron pairs in the target nucleus and
lends support for exploiting the (e,e$'$pn) to study tensor
correlations in the medium.  Globally, in super-parallel kinematics,
the induced polarization component P$_n$ is a measure for the amount
of two-nucleon knockout strength that can be attributed to the central
short-range correlations.

\begin{figure}
\begin{center}
\setlength{\unitlength}{1cm}
\begin{picture}(22,12)
\put(-1.00,-2.0){\mbox{\epsfysize=15.cm\epsffile{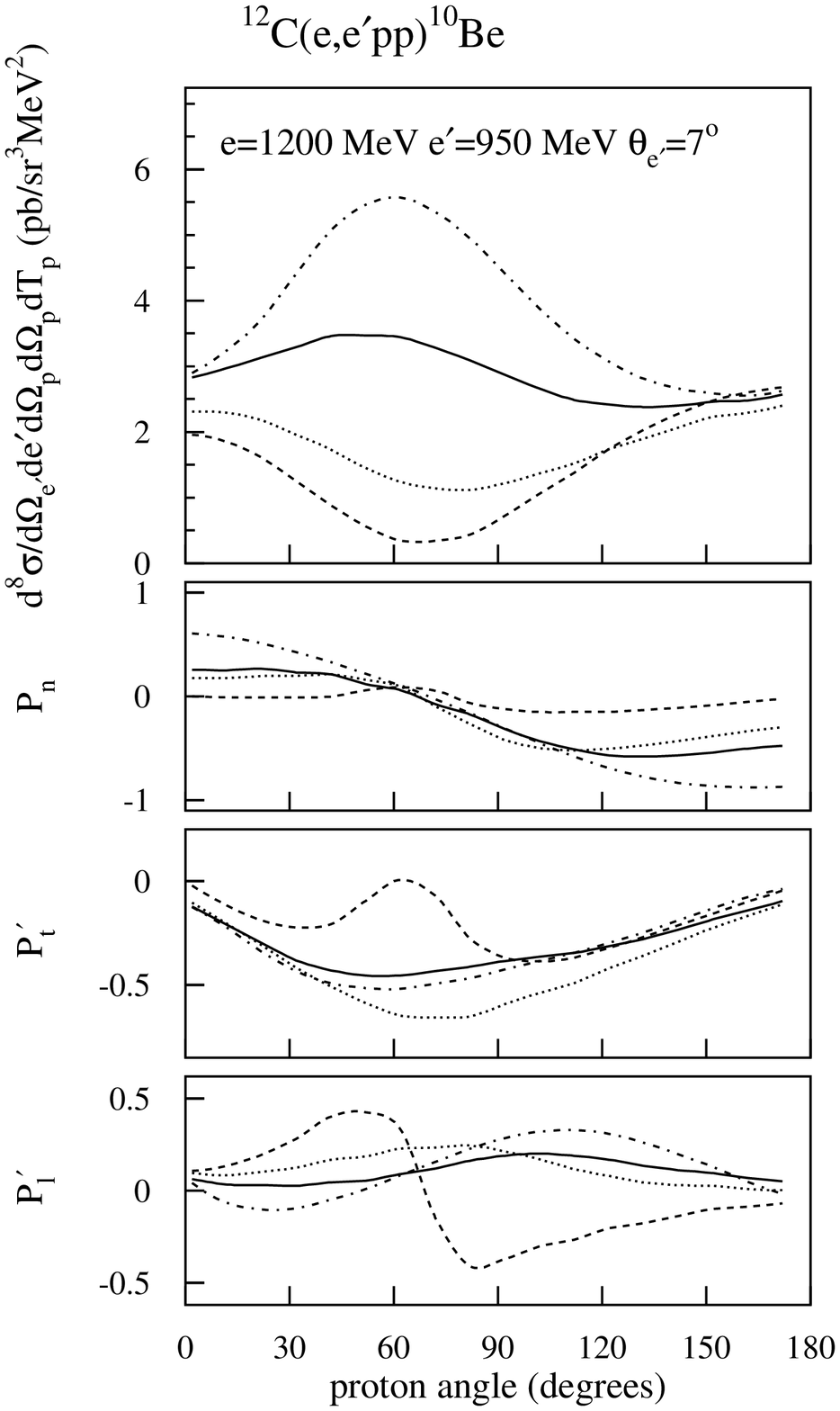}}}
\put(6.50,-2.0){\mbox{\epsfysize=15.cm\epsffile{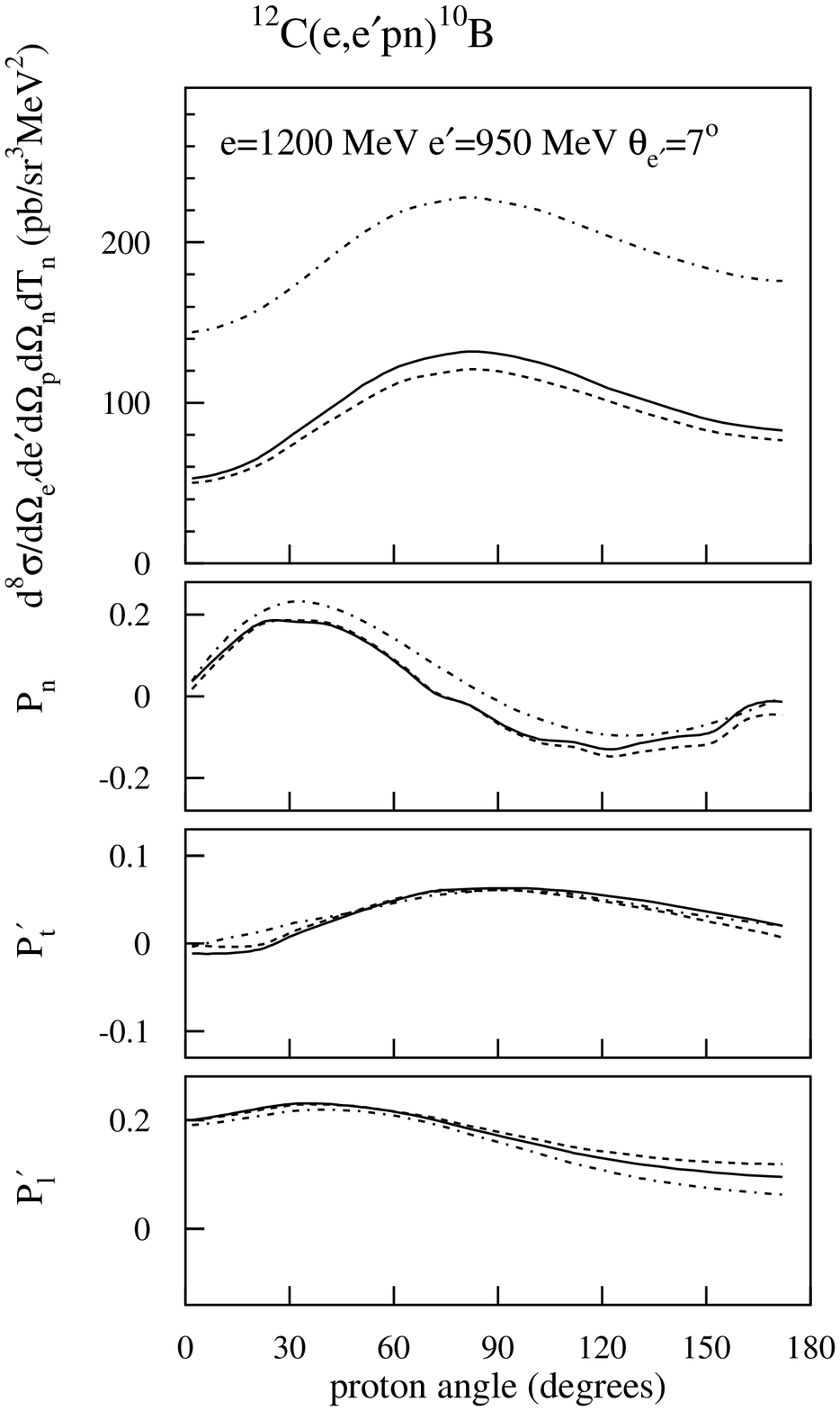}}}
\end{picture}
\end{center}
\caption{The polar angle dependence of the $^{12}$C(e,e$'$pp) and
(e,e$'$pn) differential cross section and polarization observables in
quasi-deuteron kinematics (P=0). The strength is integrated over the
missing-energy range where strength from the (1p)$^2$ shells is
expected.  The solid curves and dotted curves are calculated in the
distorted-wave approximation including the $\Delta$-current and
ground-state correlations. The solid (dotted)  curve is obtained with the
correlation function from Ref.~\protect \cite{gearhart} (\protect
\cite{pieper}).  The dot-dashed curve is the equivalent
of the solid line but is calculated with plane wave outgoing nucleons
waves.  The dashed line is the result of a distorted-wave calculation
including only the $\Delta$ current.}
\label{fig:qdres}
\end{figure}

Two-nucleon knockout studies that discriminate between the different
final states impose hefty requirements as far as experimental energy
resolution is concerned. In Fig.~\ref{fig:qdres} predictions for
proton-proton and proton-neutron knockout from $^{12}$C are shown.
For these results we have integrated the strength in the full
(p-shell)$^2$ region (which would correspond with a missing-energy
range of roughly 28 $\leq$ E$_{2m}$ $ \leq $ 40~MeV) creating
conditions that would be accessible in a moderate energy resolution
experiment.  We considered so-called quasi-deuteron (QD) kinematics
that is defined by imposing the condition $\vec{P}=\vec{0}$.  For a
given missing energy, polar angle $\theta _p$ and electron kinematics
this constraint determines the variables T$_p$,T$_n$ and $\theta_n$ in
(e,e$'$pn) and likewise for the (e,e$'$pp) case.  Note that the
$\theta _p$=0$^o$ point in QD kinematics corresponds with the $P$=0
situation in super-parallel kinematics.  Thus, for the $\theta
_p$=0$^o$ case in QD kinematics, the P$_n$ and P$'_t$ are
vanishing when solely transverse strength is contributing to the cross
section.  Comparing the (e,e$'$pn) to the (e,e$'$pp) predictions, the
latter are globally characterized by larger polarization
observables. In line with the conclusions drawn in super-parallel
kinematics, the effect of the central correlations is substantial in
the (e,e$'$pp) channel and rather small in the (e,e$'$pn) case.  Also
shown in Figure~\ref{fig:qdres} is the result of a plane wave
calculation.  As is usually the case, the P$'_t$ and P$'_l$ that
involve polarized electrons are less affected by the distortion
effects related to the final-state interaction of the ejectiles with
the residual nucleus, than the unpolarized differential cross section
and the induced polarization.

We realize that polarization measurements in triple-coincidence
reactions are a challenging task. On the other hand, we deem it to be
a valid alternative for absolute cross section measurements as the
polarization observables are predicted to be rather large.  From an
experimental point of view the induced polarization and polarization
transfer can be determined through ratios which allows one to divide
out some systematic uncertainties and bypass some tedious callibration
tasks.  Also in polarized one-nucleon knockout there are good chances
of observing indirect indications for two-nucleon knockout. Recent
measurements at MIT-Bates \cite{mit2} revealed a non-zero induced
polarization (0 $\leq$ P$_n$ $\leq$ 0.5) for the deep continuum
(missing energies larger than 50~MeV) in
$^{12}$C(e,e$'\vec{\mathrm{p}}$).  This missing-energy range is open
for two (and more) nucleon knockout.  The above results suggest that
the normal component of the induced polarization P$_n$ in
electronuclear two-nucleon knockout is quite sizeable and therefore
the measurements of Ref.~\cite{mit2} are not incompatible with a
picture in which 2N knockout mechanisms are feeding the (e,e$'$p)
reaction at high missing energies.

Concluding, we have studied the polarization observables in
electronuclear two-nucleon knockout from finite nuclei.  Particularly
in the (e,e$'$pp) case the polarization observables turned out to be
large and sensitive to the presence of scalar short-range correlations
in the ground-state of the target nucleus.  The double polarization
observable P$'_t$ has the additional advantage of being almost
insensitive to final-state interaction effects.

{\bf Acknowledgement}

This work was supported by the Fund for Scientific Research of
Flanders. Stimulating discussions with G. Rosner are gratefully
acknowledged.

\newpage

\newpage

\newpage

\begin{thebibliography}{99}
\bibitem{benhar97} 
C. Ciofi degli Atti and S. Simula, Phys. Rev. C {\bf 53} (1996) 1689
and references quoted therein.
%O. Benhar, S. Fantoni, G.I. Lykasov and
%N.N. Slavin, Phys. Rev. C {\bf 55} (1997) 244.
\bibitem{benhar94} O. Benhar, A. Fabrocini, S. Fantoni and I. Sick,
Nucl. Phys. {\bf A579} (1994) 493.
\bibitem{ciofi94} C. Ciofi degli Atti and S. Simula, Phys. Lett. {\bf
B325} (1994) 276.
\bibitem{laget}
J.M. Laget, Phys. Rev. C {\bf 35} (1987) 832.
\bibitem{blom98} 
K.I. Blomqvist {\em et al.}, Phys. Lett. {\bf B421} 
(1998) 71.
\bibitem{guenther} G. Rosner, {\em in} Proc. Conf. on Perspectives in Hadronic Physics,
ICTP Trieste, Italy, May 12-16, 1997, eds. S. Boffi, C. Ciofi degli
Atti and M.M. Gianninni (World Scientific, Singapore) 185.

P. Bartsch {\em et al.}, {\em ``Investigation of short-range 
nucleon-nucleon correlations using the reaction
$^{16}$O(e,e$'$pp)$^{14}$C in super-parallel kinematics'' } MAMI
proposal A1/1-97 (spokesperson G. Rosner), 1997.  
\bibitem{gerco} 
C.J.G. Onderwater {\em et al.}, Phys. Rev. Lett. {\bf
78} (1997) 4893 and Ph.D. thesis, Vrije Universiteit Amsterdam (1998),
unpublished.
\bibitem{raskin} A.S. Raskin and T.W. Donnelly, Ann. of Phys. {\bf
191} (1989) 78.
\bibitem{boffigiusti} S. Boffi, C. Giusti, F.D. Pacati and M. Radici,
Electromagnetic Response of Atomic Nuclei, Oxford Studies in Nuclear
Physics (Clarendon Press, Oxford, 1996).
\bibitem{donnelly} T.W. Donnelly, {\em in} Adv. in Nucl. Phys.,
eds. J.W. Negele and E. Vogt (Plenum Press, New York, 1996) 37.
\bibitem{giustifirst} C. Giusti and F.D. Pacati, Nucl. Phys. {\bf
A535} (1991) 573.
\bibitem{giusti98} G. Giusti, F.D. Pacati, K. Allaart, W. Geurts,
H. Muether and W. Dickhoff, Phys. Rev. C {\bf 57} (1998) 1691.
\bibitem{jan97} J. Ryckebusch, V. Van der Sluys, K. Heyde, H. Holvoet,
W. Van Nespen, M. Waroquier and M. Vanderhaeghen, Nucl. Phys. {\bf
A624} (1997) 581.
\bibitem{douglas} I.J.D. MacGregor {\em et al.}, Phys. Rev. Lett. {\bf
80} (1998) 245.
\bibitem{giam} Giampaolo Co$'$ and Antonio M. Lallena, Phys. Rev. C
{\bf 57} (1998) 145.
\bibitem{gearhart} C.C.Gearhart, PhD thesis, Washington University
(St. Louis, 1994), unpublished and W. Dickhoff, private communication.
\bibitem{osterfeld}
B. K\"{o}rfgen, F. Osterfeld and T. Udagawa, Phys.
Rev. C {\bf 50} (1994) 1637.
\bibitem{lee97} F.X. Lee, C. Bennhold and L.E. Wright, Phys. Rev. C
{\bf 55} (1997) 318.
\bibitem{kelly} J.J. Kelly, Adv. Nucl. Phys. {\bf 23} (1996) 75.
\bibitem{geurts} W.J.W. Geurts, K. Allaart, W.H. Dickhoff, H. M\"uther,
Phys. Rev. C {\bf 54} (1996) 1144.
\bibitem{gottfried} K. Gottfried, Nucl. Phys. {\bf 5} (1958) 557.  
\bibitem{janplb} J. Ryckebusch, Phys. Lett. {\bf B383} (1996) 1.
\bibitem{ryc98} J. Ryckebusch, D. Debruyne and W. Van Nespen,
Phys. Rev. C {\bf 57} (1998) 1319.
\bibitem{pieper} S.C. Pieper, R.B. Wiringa and V.R. Pandharipande,
Phys. Rev. C {\bf 46} (1992) 1741. 
\bibitem{mit2} R.J. Woo {\em et al.}, Phys. Rev. Lett. {\bf 80}
(1998) 456.
%\bibitem{mit1} B.D. Milbrath {\em et al.}, Phys. Rev. Lett. {\bf 80}
%(1998) 452.
\end{thebibliography}
\end{document}